\documentclass[%
reprint, 
tightenlines, 
superscriptaddress, 
showpacs, 
nobibnotes, 
amsmath, amssymb, 
aps, 
pra,
floatfix, 
dvipdfmx
]{revtex4-1}

\usepackage{latexsym}  
\usepackage{graphicx}
\usepackage{dcolumn}
\usepackage{bm}
\usepackage{ulem} 
\usepackage{multirow} 
\usepackage{booktabs}

\allowdisplaybreaks

\begin{document}


\title{Morphological Superfluid in a Nonmagnetic Spin-$2$ Bose-Einstein Condensate}

\author{Emi Yukawa} 
\email{emi.yukawa@riken.jp} 
\affiliation{RIKEN Center for Emergent Matter Science, 2-1, Hirosawa, Wako-shi, Saitama, 351-0198, Japan} 
\author{Masahito Ueda} 
\affiliation{Department of Physics, University of Tokyo, 7-3-1, Hongo, Bunkyo-ku, Tokyo, 113-8654, Japan} 
\affiliation{RIKEN Center for Emergent Matter Science, 2-1, Hirosawa, Wako-shi, Saitama, 351-0198, Japan} 

\date{\today}

\begin{abstract}
The two known mechanisms for superflow are the gradient of the U($1$) phase and the spin-orbit-gauge symmetry. 
We find the third mechanism, namely a spatial variation of the order-parameter morphology protected by a hidden su($2$) symmetry in a nonmagnetic 
spin-$2$ Bose-Einstein condensate. 
Possible experimental situations are also discussed.   
\end{abstract}

\maketitle 

Superflow is usually generated by a gradient of the U(1) phase. 
In spinor Bose-Einstein condensates (BECs), the spin-gauge symmetry provides the second mechanism of superfluidity. 
For instance, in a ferromagnetic spin-$1$ BEC, a superfluid can be induced by a spin texture via the spin-gauge symmetry~\cite{Mermin,Ho}, while a polar 
superfluid can only be carried by the gradient of the U(1) phase~\cite{Ho}. 
Similarly, in superfluid $^3$He-A phase, superflow can be induced by a texture of the $\bm{l}$-vector via the orbital-gauge symmetry~\cite{Vollhardt}. 
Here we report our finding that for the case of a spin-$2$ BEC, a spatial variation of the order-parameter shape can generate a supercurrent even in the 
nonmagnetic nematic and cyclic phases, offering the hitherto unexplored third mechanism of superfluidity. 
A full investigation of this possibility is the main theme of this Letter.

A spin-$F$ BEC can be described in the mean-field approximation by a $(2F+1)$-component order parameter 
$\bm{\psi} \equiv ({\psi}_{-F}, {\psi}_{-F+1}, \cdots , {\psi}_m, \cdots , {\psi}_{F})^T$~\cite{Ho,Ohmi}. 
The superfluid velocity is defined in terms of the order parameter, the atom mass $M$, and the local density $\rho = \sum_{m=-F}^F |{\psi}_m|^2$ 
as $\bm{v} \equiv (\hbar / 2Mi \rho ) [ {\psi}_m^* (\nabla {\psi}_m ) - (\nabla {\psi}_m^*) {\psi}_m]$.  
In a spin-$1$ BEC, the order parameter can be expressed in the irreducible representation as 
\begin{align} 
	\bm{\psi} = e^{i\varphi} \sqrt{\rho} \ R^{F=1} (\alpha ,\beta ,\gamma) \begin{pmatrix} \cos {\vartheta} \\ 0 \\ \sin {\vartheta} \end{pmatrix},  
	\label{eq:op-spin1}
\end{align} 
where $\varphi$ is the U($1$) phase, $\vartheta$ characterizes the relative amplitude between the $m = \pm 1$ states, and 
$R^F(\alpha ,\beta ,\gamma) = \exp {(-\alpha F_z)} \exp {(-\beta F_y)} \exp {(-\gamma F_z)}$ describes an Euler rotation in terms of the spin-$F$ matrices 
$F_{\mu}$'s ($\mu = x, y, z$) and the Euler angles $\alpha$, $\beta$, and $\gamma$. 
Equation.~(\ref{eq:op-spin1}) describes a ferromagnetic state at $\vartheta = n \pi /2$ ($n \in \mathbb{Z}$) and a polar state with 
$\vartheta = (2n+1) \pi /4$. 
The superfluid velocity for a spin-$1$ BEC can be expressed in terms of these parameters as~\cite{Yukawa}   
\begin{align}
	 \bm{v} = \frac{\hbar}{M} \{ (\nabla \varphi ) - [ (\nabla \alpha ) \cos {\beta} + (\nabla \gamma ) ] \cos {2\vartheta} \} . \label{eq:v-spin1}
\end{align} 
This implies that in a nonmagnetic spin-$1$ BEC a superflow can only be generated from the U($1$) phase. 

\begin{figure}[t]
	\begin{center} 
		\includegraphics[bb =  0 0 460.8 345.6, clip, width = \columnwidth]{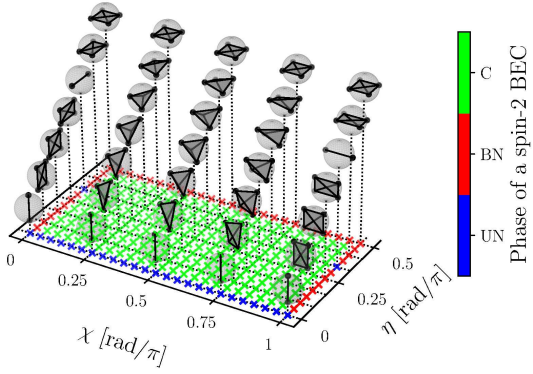}
	\end{center} 
	\caption{(Color Online) Phase diagram and stereographically mapped polygons plotted against $\chi$ and $\eta$ in Eq.~(\ref{eq:op-spin2}). 
	The blue, red, and green regions respectively show uniaxial nematic (UN), biaxial nematic (BN), and cyclic (C) phases, respectively. 
	Each polygon shows the stereographic projection of the order parameter on the Bloch sphere (see the text). } 
	\label{fig:phasediagram}
\end{figure}

However, a new situation arises for a nonmagnetic spin-$2$ BEC, where the order parameter can be parametrized by seven 
variables:  
\begin{align} 
	&\bm{\psi} 
	= e^{i\varphi} \sqrt{\frac{\rho}{2}} \ R^{F=2} (\alpha ,\beta ,\gamma ) \begin{pmatrix} 
	e^{i\chi} \sin {\eta} \\ 0 \\ \sqrt{2} \cos {\eta} \\ 0 \\ e^{i\chi} \sin {\eta} 
	\end{pmatrix}. \label{eq:op-spin2}
\end{align} 
Here, $\eta$ and $\chi$ describe the relative amplitude and phase between the $m = \pm 2$ and $m = 0$ components. 
The symmetry of this order parameter can be represented by the reciprocal spin representation~\cite{Barnett1,Turner,Barnett2} 
which is the stereographic mapping of the four roots of the following algebraic equation:  
\begin{align}
	&\sum_{m=-2}^2 \sqrt{\frac{24}{(2+m)!(2-m)!}} \ {\xi}_m^* w^{2 + m} = 0, \label{eq:rs_eqn} 
\end{align} 
where ${\xi}_m \equiv {\psi}_m / \sqrt{\rho}$ is the $m$ component of the normalized order parameter. 
In the case of a spin-$2$ BEC, Eq.~(\ref{eq:rs_eqn}) has four roots that can be stereographically mapped onto the Bloch sphere via 
$w = e^{i\phi} \tan {(\theta / 2)}$ with $\phi$ and $\theta$ being the azimuth and polar angles. 
The polygon constructed from these roots can be a line segment, a rectangle, and a tetrahedron corresponding to the uniaxial nematic, biaxial nematic, 
and cyclic phases, respectively. 
The dependence of the order parameter of a spin-$2$ BEC on $\chi$ and $\eta$ is illustrated in Fig.~\ref{fig:phasediagram}. 
Here, we note that the phase difference between the $m = \pm 2$ states can be absorbed in the Euler angle $\gamma$. 

It follows from Eq.~(\ref{eq:op-spin2}) that the superfluid velocity $\bm{v}$ is given by 
\begin{align} 
	\bm{v} = \frac{\hbar}{M} \left [ (\nabla \varphi ) + \frac{1}{2} (\nabla \chi ) (1 - \cos {2\eta}) \right ]. \label{eq:v-spin2}
\end{align} 
Remarkably, the second term, which is absent in a nonmagnetic spin-$1$ BEC, implies that a supercurrent can be generated by a texture of the 
order parameter, the physical origin of which is a spatial variation of the morphology of the order parameter. 
Taking a circulation of  Eq.~(\ref{eq:v-spin2}) along a two-dimensional closed loop $\mathcal{C} (x, y)$, we obtain 
\begin{align} 
	&\oint_{\mathcal{C}(x,y)} \left [ \frac{M}{\hbar} \bm{v} - (\nabla \varphi ) \right ] \cdot d\bm{l} \nonumber \\ 
	=& \frac{1}{2} \int_{\mathcal{S}^{\prime} (2\eta (x,y),\chi (x,y))} d(1-\cos {2\eta}) d\chi .  
	\label{eq:circ} 
\end{align}  
where $\mathcal{S}^{\prime} (2\eta (x,y),\chi (x,y))$ represents a surface of a unit sphere in spin space swept by the polar coordinates 
$(2\eta ,\chi )$ when they are mapped from the region inside the loop $\mathcal{C}$ on the $x$-$y$ plane. 
The right-hand side of Eq.~(\ref{eq:circ}) may be interpreted as one half of the Berry phase swept by the unit vector $\hat{\bm{n}}$ with the azimuth angle 
$\chi$ and the polar angle $2\eta$, which is analogous to the circulation of a supercurrent in a fully-polarized BEC~\cite{Ho} except for the coefficient 
$1/2$. 

To understand this analogy between the superfluid circulations in a nonmagnetic spin-$2$ BEC and in a ferromagnetic BEC, we let 
$\alpha = \beta = \gamma = 0$ in Eq.~(\ref{eq:op-spin2}). 
No generality is lost by this choice of the order parameter, since the Euler angles specify the direction of the order parameter but not its morphology. 
Then the nonvanishing components of the quardrupole, octupole, and hexadecapole moments are given by  
\begin{align} 
	&D_{xy} \equiv \sqrt{\frac{5}{21}} (F_x^2 - F_y^2), \label{eq:mtrx-dxy} \\ 
	&Y_{\mathrm{hyp}} \equiv \frac{\sqrt{5}}{3\sqrt{7}} (-F_x^2 - F_y^2 + 2F_z^2), \label{eq:mtrx-y} \\ 
	&T_{xyz} \equiv \frac{\sqrt{5}}{6\sqrt{3}} \overline{F_xF_yF_z}, \label{eq:mtrx-txyz} \\ 
	&{\Phi}^s \equiv \frac{\sqrt{5}}{12} (F_x^4 + F_y^4 - \overline{F_x^2F_y^2}), \label{eq:mtrx-phis} \\ 
	&{\Phi}^a \equiv \frac{\sqrt{5}}{6\sqrt{7}} (F_x^4 - F_y^4 + \overline{F_y^2F_z^2} - \overline{F_z^2F_x^2}), \label{eq:mtrx-phia} \\
	&{\Phi}^z \equiv \frac{1}{12\sqrt{7}} (3F_x^4 + 3F_y^4 + 8F_z^4 \nonumber \\ 
	&+ \overline{F_x^2F_y^2} - 4 \overline{F_y^2F_z^2} - 4 \overline{F_z^2F_x^2}), 
	\label{eq:mtrx-phiz}
\end{align} 
where the coefficients are determined so as to make the squared norm of each matrix in Eqs.~(\ref{eq:mtrx-dxy})-(\ref{eq:mtrx-phiz}) is equal to that  of 
$F_{\mu}$'s, and $\overline{F_{{\mu}_1} \cdots F_{{\mu}_n}}$ denotes the symmetrized product of $F_{{\mu}_i}$'s (${\mu}_i = x, y, z$), that is, 
$\overline{F_{{\mu}_1} \cdots F_{{\mu}_n}} = \sum_{({\nu}_1, \cdots , {\nu}_n) \in \mathrm{P}(\{{\mu}_1, \cdots , {\mu}_n \})} F_{{\nu}_1} \cdots F_{{\nu}_n}$ 
with the permutation group $\mathrm{P}(\{{\mu}_1, \cdots , {\mu}_n \})$~\cite{Shiina}. 
The physics behind the morphological supercurrent is a hidden su($2$) symmetry whose generators can be constructed from 
Eqs.~(\ref{eq:mtrx-dxy})-(\ref{eq:mtrx-phiz}) as follows: 
\begin{align} 
	&N_1 = \frac{2}{\sqrt{7}} D_{xy} - \sqrt{\frac{3}{7}} {\Phi}^a, \label{eq:mtrx-x1} \\ 
	&N_2 = - T_{xyz}, \label{eq:mtrx-x2} \\ 
	&N_3 = - \frac{2}{\sqrt{7}} Y_{\mathrm{hyp}} - \frac{1}{2} {\Phi}^s + \frac{\sqrt{5}}{2\sqrt{7}} {\Phi}^z. \label{eq:mtrx-x3}
\end{align} 
Note that the structure factor of this algebra is $2\sqrt{5}$ which is to be distinguished from the usual spin su($2$) subalgebra with the unit structure factor. 
Substituting Eq.~(\ref{eq:op-spin2}) into Eqs.~(\ref{eq:mtrx-dxy})-(\ref{eq:mtrx-phiz}), we obtain the expectation values $\langle N_i \rangle$'s of $N_i$'s
in Eqs.~(\ref{eq:mtrx-x1})-(\ref{eq:mtrx-x2}): 
\begin{align} 
	&\langle N_1 \rangle = \sqrt{5} \rho \cos {\chi} \sin {2\eta}, \\ 
	&\langle N_2 \rangle = \sqrt{5} \rho \sin {\chi} \sin {2\eta}, \\ 
	&\langle N_3 \rangle = \sqrt{5} \rho \cos {2\eta}, 
\end{align} 
which together form a vector as $\langle \bm{N} \rangle \equiv (\langle N_1 \rangle , \langle N_2 \rangle , \langle N_3 \rangle )^T$ pointing in the 
direction of $\hat{\bm{n}}$. 
In a spin-$2$ nonmagnetic BEC, $\hat{\bm{n}}$, which originates from the magnetic multipoles, plays the role of the spin vector in a fully-polarized BEC. 

A nonmagnetic superflow can also be induced between two weakly coupled BECs with different order-parameter symmetries, which we refer to as a 
morphological Josephson current. 
We assume that two nonmagnetic BECs are placed on the left and right of a high potential wall in a well localized manner. 
Then, the mean-field energy functional can be well approximated as  
\begin{align} 
	&E_{\mathrm{tot}} [{\bm{\psi}}_{\mathrm{L}}, {\bm{\psi}}_{\mathrm{R}}]= E [{\bm{\psi}}_{\mathrm{L}}] + E [{\bm{\psi}}_{\mathrm{R}}] \nonumber \\ 
	&+ K \sum_{m=-2}^2 \int d\bm{r} ({\psi}_{\mathrm{L}m}^* {\psi}_{\mathrm{R}m} 
	+ {\psi}_{\mathrm{R}m}^* {\psi}_{\mathrm{L}m}), \label{eq:enefunc_tot}
\end{align} 
where ${\bm{\psi}}_j$ ($j = \mathrm{L, R}$) represents the order parameters of the left ($\mathrm{L}$) and right ($\mathrm{R}$) BECs, 
$E [{\bm{\psi}}_j]$ indicates the energy functional of the BEC on each side, and $K$ denotes the coupling between them. 
When atoms interact via $s$-wave channels, the energy functional $E [{\bm{\psi}}_{j}]$ is given by  
\begin{align} 
	&E [{\bm{\psi}}_j] = \int d\bm{r} \biggl [ \frac{{\hbar}^2}{2M} \sum_{m=-2}^2 |(\nabla {\psi}_{jm} )|^2 + U(\bm{r}) {\rho}_{j} \nonumber \\ 
	&+ \sum_{m=-2}^2 q_j m^2 {\psi}_{jm}^* {\psi}_{jm} + \frac{1}{2} ( c_0 {\rho}_{j}^2 + c_1 {\bm{f}}_{j}^2 
	+ c_2 |A_{j}|^2 ) \biggr ], \label{eq:enefunc}
\end{align} 
where $q_j$ denotes the quadratic Zeeman energy in each well and the coupling strengths $c_0$, $c_1$, and $c_2$ are 
given by $c_0 \equiv (4{\hbar}^2 / M) (4 a_2 + 3 a_4)$, $c_1 \equiv (4{\hbar}^2 / M) (a_4 - a_2)$, and 
$c_2 \equiv (4{\hbar}^2 / M) (7 a_0 - 10 a_2 + 3a_4)$. 
Here, $a_{\mathcal{F}}$'s represent the scattering lengths for binary collisions with the total hyperfine spins $\mathcal{F} = 0$, $2$, and $4$. 
The density, the magnetization vector, and the spin-singlet amplitude are denoted by ${\rho}_j$, 
${\bm{f}}_j \equiv \sum_{m,n=-2}^2 (\bm{F})_{mn} {\psi}_{jm}^* {\psi}_{jn}$, and $A_{j} \equiv \sum_{m,n=-2}^2 (A)_{mn} {\psi}_{jm} {\psi}_{jn}$ with 
a five-by-five anti-diagonal matrix $(A)_{mn} \equiv \mathrm{adiag} (1, -1, 1, -1, 1) / \sqrt{5}$. 
Then, the multicomponent Gross-Pitaevskii equation for ${\bm{\psi}}_{j}$ can be obtained from Eq.~(\ref{eq:enefunc_tot}) as 
$i\hbar (d{\psi}_{jm}/dt) = \delta E_{\mathrm{tot}}/\delta {\psi}_{jm}^*$, from which we obtain $d {\rho}_{\mathrm{L}} /dt= - d{\rho}_{\mathrm{R}} /dt$ and 
\begin{align}
	\frac{d {\rho}_{\mathrm{L}}}{dt} = \frac{K}{i\hbar} \sum_{m=-2}^2 ({\psi}_{\mathrm{L}m}^* {\psi}_{\mathrm{R}m} 
	- {\psi}_{\mathrm{R}m}^* {\psi}_{\mathrm{L}m}). \label{eq:rho_L}
\end{align} 

To derive a general non-magnetic current-phase relation, let us take the initial order parameter in Eq.~(\ref{eq:op-spin2}) with 
${\alpha}_j = {\beta}_j = {\gamma}_j = 0$ and assume that ${\bm{\psi}}_j$ is uniform in each well and exponentially decays on the other side of the potential 
wall. 
Then, the populations of the $m = \pm 1$ components stay zero and those of the $m = \pm 2$ components remain equal to each other, since no population 
transfer occurs between $m = \pm 2, 0$ and $\pm 1$ and the energy functional in Eq.~(\ref{eq:enefunc_tot}) is symmetric with respect to exchange of the 
$m = \pm 2$ states, from which we conclude that the order parameter can be expressed as in Eq.~(\ref{eq:op-spin2}) with 
${\alpha}_j = {\beta}_j = {\gamma}_j = 0$ during the time evolution. 
Then, Eq.~(\ref{eq:rho_L}) reduces to   
\begin{align} 
	\frac{d {\rho}_{\mathrm{L}}}{dt} = &\frac{2K}{\hbar} \sqrt{{\rho}_{\mathrm{L}} {\rho}_{\mathrm{R}}} 
	[ \sin {\Delta \varphi} ( \cos {\Delta \chi} \sin {{\eta}_{\mathrm{L}}} \sin {{\eta}_{\mathrm{R}}} \nonumber \\ 
	&+ \cos {{\eta}_{\mathrm{L}}} \cos {{\eta}_{\mathrm{R}}} ) + \cos {\Delta \varphi} \sin {\Delta \chi} 
	\sin {{\eta}_{\mathrm{L}}} \sin {{\eta}_{\mathrm{R}}} ], \label{eq:rho_L2}
\end{align} 
where $\Delta \varphi \equiv {\varphi}_{\mathrm{R}} - {\varphi}_{\mathrm{L}}$ and $\Delta \chi \equiv {\chi}_{\mathrm{R}} - {\chi}_{\mathrm{L}}$. 
When the left BEC is in the biaxial nematic phase $({\varphi}_{\mathrm{L}} = 0, {\chi}_{\mathrm{L}} = 0, {\eta}_{\mathrm{L}} = \pi / 2)$ 
and the right BEC is in the cyclic phase 
$({\varphi}_{\mathrm{R}} = \Delta \varphi , {\chi}_{\mathrm{R}} = \Delta \chi \neq 0, {\eta}_{\mathrm{R}} = \pi / 4)$, Eq.~(\ref{eq:rho_L2}) gives 
\begin{align}
	\frac{d {\rho}_{\mathrm{L}}}{dt} = \frac{\sqrt{2} K}{\hbar} \sqrt{{\rho}_L {\rho}_R} \ \sin {(\Delta \varphi + \Delta \chi)}, 
\end{align} 
which implies that the current flows depending on the differences in the parameter $\chi$ determining the shape of the order parameter 
and the U($1$) gauge $\varphi$. 
Thus the supercurrent flows in a manner depending on the difference in the morphology of the order parameter between the left and right BECs. This is 
essentially different from the Josephson effect due to the Goldstone modes associated with symmetry breaking from 
$O(N)$ to $O(N-1)$~\cite{Smerzi,Leggett,Qi,Esposito}.
When the two BECs share the same morphology, Eq.~(\ref{eq:rho_L2}) reduces to the familiar Josephson relation caused by the difference in the U($1$) 
phase. 

\begin{figure*}[t]
	\begin{center} 
		\includegraphics[bb = 0 0 720 720, clip, width = \textwidth]{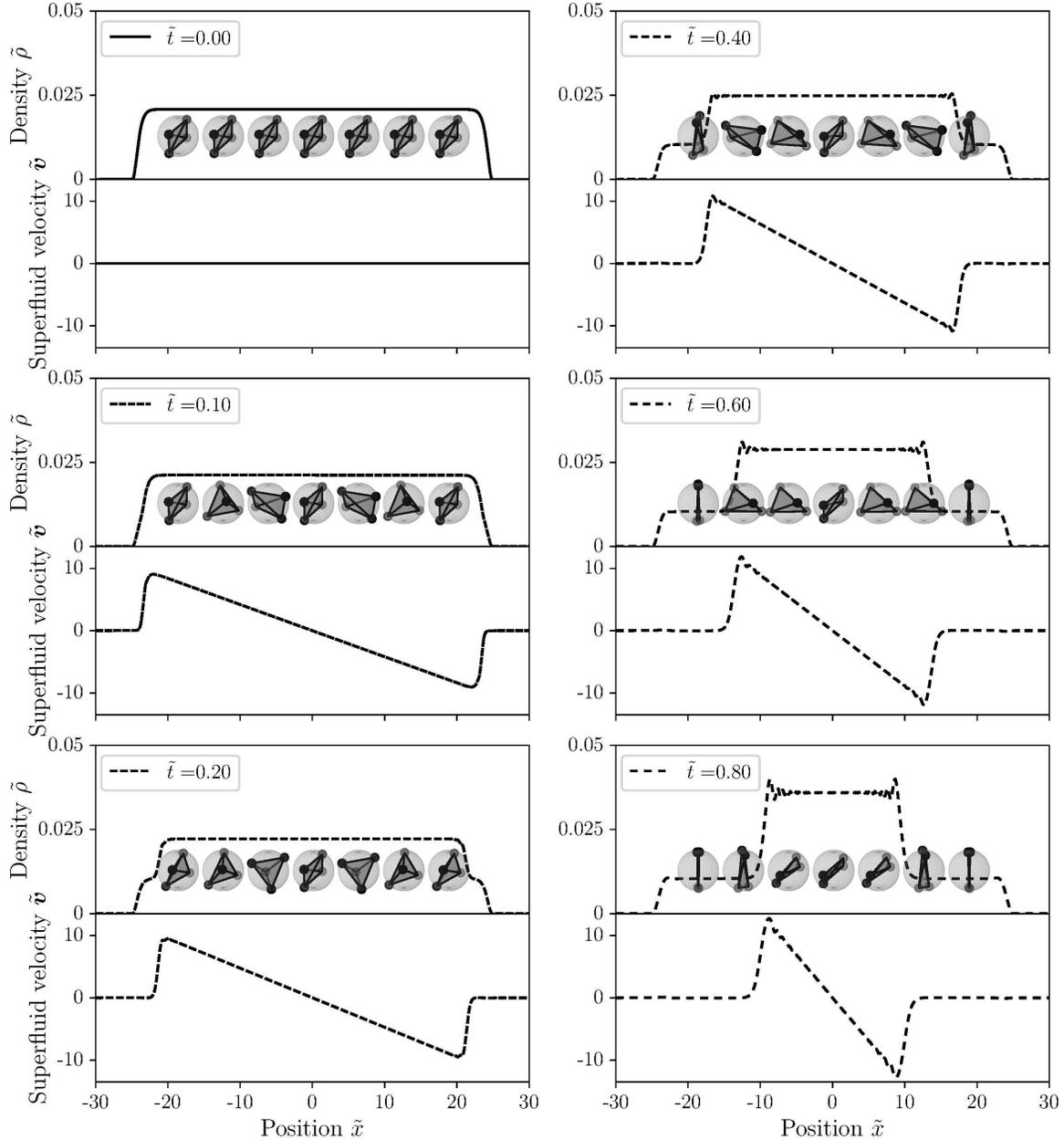}
	\end{center} 
	\caption{Position dependences of the density profile and the morphology of the order parameter (upper panels) and 
	the superfluid velocity (lower panels) of a spin-$2$ BEC at $\tilde{t} \equiv \omega_x t = 0$, $0.1$, $0.2$, $0.4$, $0.6$ and $0.8$. 
	The quadratic Zeeman field is switched on during $\tilde{t} = 0$-$0.1$. 
	The coordinate, the density, and the superfluid velocity are normalized as 
	$\tilde{x} \equiv x / l_x$, $\int d\tilde{x} \tilde{\rho} (\tilde{t}, \tilde{x}) =1$, and $\tilde{v} = v / l_x {\omega}_x$. } 
	\label{fig:dens_and_v_prof}
\end{figure*} 
We now demonstrate the above general theory by numerical simulation. 
The nonmagnetic supercurrent given in Eq.~(\ref{eq:v-spin2}) can be induced by a spatially dependent quadratic Zeeman effect. 
To demonstrate this, we consider a cigar-shaped spin-$2$ BEC of $10^3$ $^{87}$Rb atoms, apply a spatially dependent quadratic Zeeman field, and 
examine how the density profile of the BEC changes after the quadratic Zeeman field is switched off. 
We assume that the axial trapping frequency ${\omega}_x = 2 \pi \times 10 \ [\mathrm{Hz}]$ in the $x$ direction is much smaller than those in the radial 
directions, i.e., ${\omega}_x \ll {\omega}_y , {\omega}_z = 2 \pi \times 200 \ [\mathrm{Hz}]$ whose ratio 
$\gamma \equiv \sqrt{{\omega}_y {\omega}_z} / {\omega}_x = 20$ characterizes the dynamics of the system~\cite{Bao}. 
Then the mean-field dynamics of the spin-$2$ BEC can be described by the following multi-component Gross-Pitaevskii equation~\cite{Bao,Ciobanu}:  
\begin{align}
	&i\hbar \frac{\partial {\psi}_m}{\partial t} = \biggl [ - \frac{{\hbar}^2}{2M} {\nabla}^2 + U(x) + q(t, x) m^2 \biggr ] {\psi}_m \nonumber \\ 
	&+ \frac{\gamma}{2\pi} \sum_{n=-2}^2 \{ [ c_0 \rho (t, x) {\delta}_{mn} + c_1 (\bm{f}(t,x) \cdot \bm{F})_{mn} ] {\psi}_n \nonumber \\
	&+ c_2 A (t, x) (A)_{mn} {\psi}_n^* \} , \label{eq:GP}
\end{align} 
where $M$ represents the mass of an $^{87}$Rb atom. 
The trapping potential $U(x)$ is assumed to be a box potential in the $x$ direction given by 
\begin{align}
	U(x) = \begin{cases} \infty & (|x| > L/2); \\ 
	 0 & (|x| \leq L/2), \end{cases} \label{eq:U}
\end{align} 
where $L = 50 \times l_x$ with $l_x = \sqrt{\hbar / M {\omega}_x} \approx 3.41 \ [\mu \mathrm{m}]$. 
In the numerical calculation, we set the height of the trapping potential to be $10^2 \times \gamma c_0/ 2\pi$. 
We vary the quadratic Zeeman field $q(t, x)$ as  
\begin{align} 
	q(t, \bm{r}) = \begin{cases} q^{\prime} x^2 & (0 \leq t < T); \\ 
	0 & (t<0 \text{ and } t \geq T), \end{cases} \label{eq:q}
\end{align}  
where $q^{\prime} = 10 h$ and $T=0.1/\omega_x$. 
The scattering lengths $a_{\mathcal{F}}$'s for binary $s$-wave collisions with their total hyperfine spins $\mathcal{F} = 0, 2$, and $4$ are given by 
$a_0 = 89.4 a_{\mathrm{B}}$, $a_2 = 94.5 a_{\mathrm{B}}$, and $a_4 = 106 a_{\mathrm{B}}$ with $a_{\mathrm{B}}$ being the Bohr radius~\cite{Ciobanu}. 
The density profile $\rho (0, x)$ of the initial order parameter is chosen to be the ground state of a scalar BEC with the same potential $U(x)$ and the 
interaction energy $c_0$ is chosen to be the same as that used in Eq.~(\ref{eq:GP}). 
The initial spin configuration is assumed to be spatially uniform and given by 
$\bm{\xi} (t=0, x) = (1, 0, \sqrt{2}, 0, 1)^T/2$ corresponding to the biaxial nematic state as shown in Fig.~\ref{fig:phasediagram}.  
By numerically solving Eq.~(\ref{eq:GP}) via the Crank-Nicolson method, the multi-component order parameter $\bm{\psi}$ can be obtained and the 
dynamics of the density profile $\rho (t, x)$, the superfluid velocity $\bm{v} (t, x)$, and the magnetization vector $\bm{f} (t, x)$ can be calculated from 
$\bm{\psi}$. 
The density profile and the superfluid velocity evolves in time as shown in Fig.~\ref{fig:dens_and_v_prof}, where 
$\bm{f} (t, x) = \bm{0}$, which implies that the BEC stays nonmagnetic throughout the time evolution. 
We also calculate the time evolution of the order-parameter morphology. 
As shown in Fig.~\ref{fig:dens_and_v_prof}, the texture of the order-parameter morphology. 

In summary, we have found the third mechanism of supercurrent that originates from a spatial variation of the morphology of the order parameter in 
nonmagnetic spin-2 BECs.
We also discuss the morphological Josephson current. 
The morphological superflow can be generated by using a spatially dependent quadratic Zeeman effect. 

This work was supported by KAKENHI Grant No. JP18H01145 and 
a Grant-in-Aid for Scientific Research on Innovative Areas ``Topological Materials Science’’ (KAKENHI Grant No. JP15H05855) from the Japan Society for 
the Promotion of Science.

\end{document}